\documentstyle[prb,aps,psfig]{revtex}
\begin{document}
\draft
%%%%%%%%%%%%%%%%%%%%%
%
%\preprint{\footnotesize Revised \today}
\twocolumn[
\title{%\vspace{-40pt}{\normalsize\null\hspace{5.5in}\sf DRAFT}\vspace{30pt}\\
Neutron diffraction study of stripe order in La$_2$NiO$_{4+\delta}$ with
$\delta=\frac2{15}$}
\author{P. Wochner and J. M. Tranquada}
\address{Physics Department, Brookhaven National Laboratory, Upton, New York
11973}
\author{D. J. Buttrey}
\address{Department of Chemical Engineering, University of Delaware, Newark,
Delaware 19716}
\author{V. Sachan$^\ast$}
\address{Materials Science Program, University of Delaware, Newark,
Delaware 19716}
\date{June 16, 1997}
\maketitle
\widetext
\advance\leftskip by 57pt
\advance\rightskip by 57pt

\begin{abstract}

We report a detailed neutron scattering study of the ordering of spins and
holes in oxygen-doped La$_2$NiO$_{4.133}$.  The single-crystal sample exhibits
the same oxygen-interstitial order but better defined charge-stripe order
than that studied previously in crystals with $\delta=0.125$.  In particular,
charge order is observed up to a temperature at least twice that of the magnetic
transition, $T_m=110.5$~K.  On cooling through $T_m$, the wave vector
$\epsilon$, equal to half the charge-stripe density within an NiO$_2$ layer,
jumps discontinuously from $\frac13$ to 0.2944.  It continues to decrease with
further cooling, showing several lock-in transitions on the way down to low
temperature.  To explain the observed lock-ins, a model is proposed in which
each charge stripe is centered on either a row of Ni or a row of O ions.  The
model is shown to be consistent with the $l$-dependence of the magnetic peak
intensities and with the relative intensities of the higher-order magnetic
satellites.  Analysis of the latter also provides evidence that the magnetic
domain walls (charge stripes) are relatively narrow.  In combination with a
recent study of magnetic-field-induced effects, we find that the charge
stripes are all O-centered at $T>T_m$, with a shift towards Ni centering at 
$T<T_m$.  Inferences concerning the competing interactions responsible for the
the temperature dependence of $\epsilon$ and the localization of charge within
the stripes are discussed.

\end{abstract}
]

\narrowtext

\section{ Introduction }

La$_2$NiO$_4$ is a Mott-Hubbard insulator consisting of antiferromagnetic
NiO$_2$ planes alternating with La$_2$O$_2$ layers.  The NiO$_2$ planes can be
doped with holes; however, contrary to conventional expectations, the material
remains nonmetallic up to quite large hole
concentrations.\cite{gopa77,cava91,stra91}  There is now considerable evidence
that the insulating behavior occurs because the dopant-induced holes tend to
order themselves in periodically spaced
stripes.\cite{chen93,cheo94,tran94a,sach95,tran95b,tran96a,naka97,lee97}  These
charge stripes run diagonally relative to the square lattice defined by the
Ni--O--Ni bonds.  In the essentially undoped regions between the stripes the Ni
spins can order antiferromagnetically, with the charge stripes acting as
antiphase domain 
walls.\cite{tran94a,sach95,tran95b,tran96a,naka97,lee97,hayd92,brow92}  
This behavior represents an excellent example of ``topological''
doping in a correlated insulator.\cite{kive96}

In comparing results obtained on samples doped both by Sr substitution and by
addition of excess oxygen, several trends have become clear.  In samples where
both charge and spin orderings have been observed, the charge generally orders
at a higher temperature than the spins.\cite{sach95,tran96a,lee97}  (The one
apparent exception is the case\cite{tran94a} of La$_2$NiO$_{4+\delta}$ with
$\delta=0.125$.  It will be shown in the present paper that, in this
composition as well, the charge orders first.)  Furthermore, both the charge and
spin ordering temperatures increase approximately linearly with the net hole
concentration $n_h$.\cite{sach95,tran96a,naka97}  Also varying linearly with
$n_h$ is the low-temperature spacing between charge stripes, corresponding to a
hole density of roughly one per Ni site along a stripe.

The observed low-temperature hole density within the stripes is consistent with
the mean-field calculations of Zaanen and Littlewood.\cite{zaan94}  They
evaluated an appropriate three-band Hubbard model at zero temperature and found
that electron-phonon couplings tend to reinforce the stability of charged
domain walls in an antiferromagnetic background.  The numerical calculations
yielded a relatively narrow charge stripe centered on a diagonal row of Ni
atoms.  Zaanen and Littlewood pointed out that the domain wall states form a
half-filled one-dimensional (1D) band, which is unstable to a Peierls
distortion.  Such a distortion would create a gap at the Fermi level, thus
explaining the insulating behavior.

Despite the apparent success of the Hartree-Fock calculations\cite{zaan94} in
describing the low-temperature stripe structure, such an approach runs into
difficulties at high temperature.  In the absence of stripe order, a
Hartree-Fock analysis of the doped nickelate would yield a metallic state,
whereas numerous studies\cite{bi90,ido91,bi93,cran93,calv96,kats96} have shown
that even above the charge-ordering temperature the optical conductivity is
dominated by midinfrared absorption bands, with no $\omega=0$ Drude component.
Furthermore, neutron diffraction studies of the ordered state have shown that
in some cases the stripe spacing changes with
temperature,\cite{tran94a,tran96a} implying a variation in the charge density
within a stripe.  Thus, one cannot rely on a unique stripe charge density to
explain the insulating behavior.

It is clear that there are mysteries remaining to be unraveled.  For example,
what is the nature of the temperature-dependent variation of the stripe
spacing?  How narrow are the charge stripes, and to what extent are they pinned
to the lattice?  Are charge stripes actually centered on Ni rows, or are other
configurations possible?  To address these issues, a well-ordered sample is
required.

Much of the work cited above has focussed on Sr-doped La$_2$NiO$_4$.  That
system has the advantage of allowing a continuous variation of the hole
concentration by dopants whose random positions are fixed at relatively high
temperature.  The down side is that the disorder of the dopants appears to
prevent the development of long-range stripe 
order.\cite{chen93,sach95,tran96a,lee97}  The alternative is to study
oxygen-doped La$_2$NiO$_{4+\delta}$.  The constraint here is that one must work
with the phases that nature provides: ordered phases occur only at special
values of $\delta$,  and for an arbitrary value of $\delta$ one may observe
phase
separation.\cite{tran95b,naka97,rodr91,hoso92,rice93,tamu93,tran94b,yazd94} 
On the positive side, long-range order of interstitials and charge stripes
coincide in a composition originally characterized as
$\delta=0.125$.\cite{tran94a,tran95b,naka97,yama94}

Analysis of neutron diffraction measurements on one particular $\delta=0.125$
crystal\cite{tran95b} indicated that the ordered interstitial structure actually
corresponds to an ideal oxygen excess of $\delta=\frac2{15}=0.133$.  Thus, the
$\delta=0.125$ composition is oxygen deficient relative to the ideal phase.  As
will be explained below, we have prepared a new crystal that appears to
correspond much more closely to the ideal value of $\delta$.  The magnetic
Bragg peaks measured on this crystal are sharper than those observed with the
$\delta=0.125$ crystal,\cite{tran94a,tran95b} thus allowing better sensitivity
to the features of interest.  We have recently reported results on
magnetic-field-induced staggered magnetization\cite{tran97b} and on spin
excitations in the stripe-ordered phase\cite{tran97c} measured on the same
crystal.  In the present paper we concentrate on various aspects of the static
stripe order in zero field.

In earlier studies\cite{tran94a,yama94} of $\delta=0.125$ crystals, an apparently
discontinuous jump from zero to finite intensity of both the magnetic and
charge-order superlattice peaks was observed on cooling below $T_m=110.5$~K. 
With the $\delta=0.133$ crystal we observe a similar discontinuous jump in the
intensities, but we find that charge order clearly survives at $T>T_m$ with a
jump in the stripe spacing to a value commensurate with a second harmonic of the
interstitial order.  In the latter regime the charge-order peak intensity
decreases exponentially with temperature, so that no clear transition to the
disordered state has been observed. 

To obtain a more precise measure of the incommensurate splitting $\epsilon$ of
the magnetic peaks at $T<T_m$, we measured the temperature dependence of the
position and intensity of a third harmonic peak.  Where previous
work\cite{tran94a} had provided weak evidence that $\epsilon$, which varies with
temperature, tends to show lock-in behavior at certain rational fractions, the
present results show clear plateaus at $\epsilon=\frac27$ (98~K$\le T\le$102~K)
and $\epsilon=\frac5{18}$ (75~K$\le T\le$92~K).  The lock-ins demonstrate that
the charge stripes are coupled to the lattice.  To explain the continuous
distribution of $\epsilon$ values found at other temperatures, an analogy is
made with the problem of finding the ground state arrangement of integral
charges on a one-dimensional lattice at arbitrary filling.  From the
solution\cite{hubb78,bak82} to the latter problem, it appears that a devil's
staircase of ordered phases should be possible; however, due to entropy,
long-range order is observed only for phases with relatively short periods. 
The fact that $\epsilon$ varies with temperature indicates that competing
interactions must be involved in determining the order.\cite{tran96a}

Coupling to the lattice is also evident in the way that the stripes align
themselves from one layer to the next.  If there were no interaction with the
lattice (as was assumed previously\cite{tran95b}), then the stripes within a
layer would be uniformly spaced, and their positions would be staggered from
one layer to the next in order to minimize the Coulomb energy.  Such a
symmetric stacking of the layers of stripes can lead to forbidden superlattice
peaks corresponding to the charge order, similar to the situation for nuclear
reflections due to the average body-centered stacking of the NiO$_2$ layers. 
However, the pinning of the charge stripes to the lattice means that the shift
of the stripe pattern from one layer to the next can only occur in increments
of the lattice spacing.  As a consequence, the intensities of the magnetic peaks
oscillate as a function of the component of the momentum transfer that is
perpendicular to the planes.  We show that the modulation of the intensities
can be modeled quantitatively in terms of the stacking structure factor when a
small amount of stacking disorder is taken into account.  It
follows that the missing charge-order peaks are not forbidden, but just
extremely weak.

To characterize the widths of the charge stripes, we have measured the higher
harmonic magnetic Bragg peaks.  Narrow domain walls give squared-off magnetic
domains, and deviations from a sinusoidal modulation are reflected in finite
Fourier components beyond the first harmonic.  Another contribution to the
higher harmonics can come from an alternation of the stripe spacing associated
with the pinning to the lattice.  We show that although the intensities of the
higher harmonics are all quite weak, with the strongest (the third) being
$<2\%$ of the first, they are consistent with narrow charge stripes. 
Furthermore, the weakness of the harmonics in the case of $\epsilon=\frac27$
indicates that stripes can be centered both on rows of Ni atoms and on rows of
O atoms.  Assuming that the stripes are Ni-centered at low temperature, the
temperature dependence of $\epsilon$ implies that the stripes become
O-centered at high temperature.  This leads to a prediction of ferrimagnetic
correlations in the stripe-ordered regime at $T>T_m$, which was recently
confirmed by measurements in a magnetic field.\cite{tran97b}

The rest of the paper is organized as follows. The next section contains a 
description of experimental procedures, and it is followed by a brief
review of nomenclature and notation.  The temperature dependences of the
intensities and positions of superlattice peaks are presented and analyzed
in Sec.~IV.  The results relevant to interlayer stacking of stripes are
treated in Sec.~V.  In Sec.~VI the intensities of the magnetic higher
harmonic satellites are analyzed.  The significance of these results is
discussed in Sec.~VII, and a brief summary is given in Sec.~VIII.

\section{ Experimental Procedures }

The single-crystal sample used in the present study was grown by radio
frequency  skull melting as described elsewhere.\cite{rice93} The oxygen
concentration was  selected by annealing at $464^\circ$~C for 5 days in flowing
O$_2$ (1~bar) followed by a quench to room temperature. The annealing
conditions were chosen to give a nominal $\delta$ of 0.150, based on
earlier work on the phase diagram.\cite{rice93} An initial neutron diffraction 
study of this crystal in 1993 indicated that the dominant oxygen-ordered phase
is the same as that found previously
\cite{tran95b} in a crystal with $\delta=0.125$. There were also  weak
diffraction peaks indicative of a second phase with a different $\delta$;
however, these were absent $\sim1.5$ years later when the present study was
initiated.  (One fact that might be relevant to the disappearance of the
second phase is that the crystal was stored in a He atmosphere, which is
relatively reducing, rather than in air.)  The room temperature lattice
parameters obtained at that later time are $a=5.461~\mbox{\AA}$  and
$c=12.674~\mbox{\AA}$. Making use of the  calibration established by the
earlier X-ray diffraction study,\cite{rice93} the $c/a$ ratio of  2.3208
indicates $\delta=0.136\pm0.005$, consistent with the ideal value of
$\frac2{15}$ for the  model ordered interstitial structure identified
\cite{tran95b} in the case of $\delta=0.125$.

The neutron diffraction measurements were performed on the H4M, H8 and H9A
triple-axis spectrometers at the High Flux Beam Reactor located at Brookhaven
National Laboratory. At H9A, which is on the cold source, neutrons of energy
5.0~meV were selected.  Effective horizontal collimations of
$40'$-$40'$-$20'$-$40'$ (from reactor to detector)  were used together with a
cold Be filter to elliminate neutrons at shorter harmonic  wavelengths. At H4M
and H8 a neutron energy of 14.7~meV, horizontal collimations of 
$40'$-$40'$-$40'$-$40'$ or $40'$-$40'$-$80'$-$80'$ and a pyrolytic graphite (PG)
filter  were used. At all spectrometers the neutrons were monochromatized and
analyzed by PG  crystals set for the (002) reflection. 

The cold  neutrons at H9A were used to study the weak higher-order satellites of
the magnetic  scattering peaks, which are strongest at small momentum transfer
${\bf Q}$. The  excellent signal to background ratio of H9A together with the
larger scattering  angles in the case of cold neutrons compared to higher energy
neutrons enabled the  determination of third-order satellite positions with high
precision. The higher  energy neutrons at H4M and H8 were required to measure
the nuclear scattering  associated with lattice displacements due to charge
order.  Except for the third-harmonic measurements, where scans were taken
along $Q=(h,0,1)$, integrated  intensities  were determined from
$\theta-2\theta$ scans.

All measurements were  performed with the crystal oriented so that the $(h,0,l)$
zone of reciprocal space  was in the horizontal scattering plane.  (The
notation and choice of unit cell are explained in the following section.)  The
aligned crystal was mounted in an  Al can filled with He exchange gas that was
attached to the cold finger of a Displex  closed-cycle refrigerator. The
temperature was measured with a Si diode and a  Pt resistance thermometer to an
accuracy of better than $\pm1$~K.   

The ordering kinetics of the oxygen
interstitials can be rather slow, and the cooling rate through the ordering
transition can affect whether long-range order is achieved.  For example,
extremely slow cooling is required to optimize the order in the
case\cite{lore95} of $\delta=0.105$, and rapid cooling also had measurable
effects in studies\cite{tran95b,naka97} of $\delta=0.125$.  In the present
case, no significant effects of cooling rate on interstitial order were
observed.  On the other hand, the
cooling and heating history for $T\le T_m$ has an influence on the relative
fractions of coexisting magnetic phases, the width of the temperature regions 
where the magnetic wave vector is locked-in to rational fractions, and the 
detailed spectral weight of higher order magnetic satellites and their peak
widths. Therefore, the temperature dependence was studied during slow,
careful warming after an initial cool down to 10~K.    

\section{ Notation } \label{Sec:Not}

In order to describe and explain the data it is first necessary to review the
structure\cite{tran95b} and explain our notation.  Within the resolution of our
instrument, the fundamental reflections of this crystal indicate a tetragonal
(HTT) symmetry (space group $I4/mmm$) down to at least 10~K, whereas the
structural superlattice peaks suggest an orthorhombic  symmetry. To describe
the diffraction peaks we will use an indexing based on an orthorhombic cell  
(or pseudo-tetragonal $F4/mmm$) with a unit cell size $\sqrt 2 a_t \times \sqrt
2 a_t \times c_t$  relative to the basic body-centered tetragonal cell.  Note
that in the body-centered tetragonal cell the $a$ and $b$ axes are parallel to
the Ni-O bonds within the NiO$_2$ planes, whereas in the larger cell that we
use the axes are rotated from the bonding directions by $45^\circ$.

Four different types of
superlattice peaks can be found in this sample. Two of them correspond to  the
ordering of the oxygen interstitials and the other two involve ordering of the
doped holes  and the Ni spins. Each set of peaks is characterized by a wave
vector  ${\bf g}_j$ such that 
${\bf Q}_{SL}$, the position of a superlattice peak of type $j$, is given by 
\begin {equation} 
  {\bf Q}_{SL}={\bf G}\pm {\bf g}_j,
  \label{eq:qsl}
\end {equation} 
where $\bf G$ is a reciprocal lattice vector corresponding to
the average pseudo-tetragonal unit cell. The oxygen order is characterized by
two wave vectors,
\begin {equation} 
  {\bf g}_{\rm O1}=\left(\frac13,0,1\right) 
\end {equation} 
and
\begin {equation} 
  {\bf g}_{\rm O2}=\left(0,\frac45,\frac45\right),
\end {equation} 
where wave vector components are specified in reciprocal lattice
units of $(2\pi /a, 2\pi /a, 2\pi /c)$.  The twinning associated with the
interstitial order is described in Ref.~\onlinecite{tran95b}.

To describe the superlattice peaks associated with stripe order it is
necessary to expand the set of reciprocal lattice vectors considered in
Eq.~(\ref{eq:qsl}).  The symmetry relationship between the two NiO$_2$ planes
per $F4/mmm$ unit cell leads to forbidden reflections.  For example, for
reciprocal lattice vectors of the type $(H,0,L)$ one has the requirements that
both $H$ and $L$ be even integers.  We will show in Sec.~V that the stripe
patterns in neighboring layers are {\it not} related by symmetry, and as a
result the relevant reciprocal lattice vectors include the case $L$ odd.

\begin{figure}
\centerline{\psfig{figure=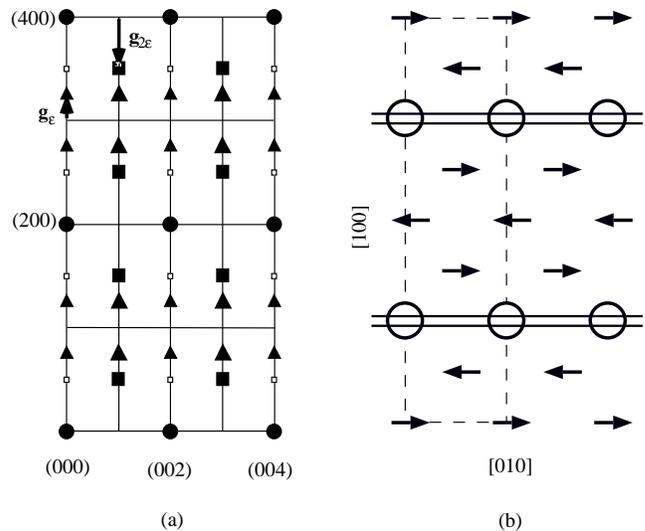,width=3.4in,bbllx=15bp,bblly=200bp,%
bburx=560bp,bbury=680bp}}
\caption{(a) Diagram of the $(h0l)$ zone, showing positions of diffraction 
peaks, with indexing based on space group F4/mmm for the average nuclear 
structure. Solid circles, fundamental Bragg peaks; solid triangles, magnetic
superlattice peaks; solid squares, nuclear superlattice peaks corresponding to 
charge order; open squares, allowed but unobserved charge-order peak positions.
Oxygen ordering peaks are excluded. (b) Model of spin and hole ordering in a
NiO$_2$ plane, as discusssed in the text; circles indicate positions of holes
(Ni centered), arrows indicate correlated Ni moments. The dashed line outlines a
unit cell, and double lines mark positions of domain walls.}
\label{fg:recip_map}
\end{figure}

The positions of the magnetic and charge-order superlattice peaks within the
$(h,0,l)$ zone are illustrated in Fig.~1(a).  The charge order is determined by 
\begin {equation} 
  {\bf g}_{2\epsilon}=(2\epsilon,0,0), 
\end {equation} 
with the average distance between domain walls in real space
equal to $a/2\epsilon$. The magnetic order involves two modulation wave
vectors.  Simple antiferromagnetic order would be described by 
${\bf g} = {\bf Q}_{\rm AF} = (1,0,0)$.  Stripe order results in a second
modulation 
\begin {equation} 
  {\bf g}_{\epsilon}=(\epsilon,0,0), 
\end {equation}
so that the net wave vector is 
${\bf g} = {\bf Q}_{\rm AF}\pm{\bf g}_\epsilon$.  In real space the spin
structure consists of locally  antiferromagnetic order with an overall
modulation period $a/\epsilon$, twice that of the charge period, reflecting
the fact that the hole stripes act as antiphase boundaries for the spin
structure.  The  idealized real-space spin and charge structure for
$\epsilon=\frac14$ is presented in  Fig.~\ref{fg:recip_map}(b).

Of course, ${\bf g}_\epsilon$ and ${\bf g}_{2\epsilon}$ represent only the
dominant sinusoidal components of the magnetic and charge-order modulations,
respectively.  Deviations from sinusoidal modulations will result in 
superlattice peaks corresponding to higher Fourier components of the order. 
The modulation wave vector for the $n^{\rm th}$ harmonic is simply
\begin {equation} 
  {\bf g}_{n\epsilon}=(n\epsilon,0,0).
\end {equation}
For the magnetic scattering, only harmonics with $n$ odd will appear, split
about ${\bf Q}_{\rm AF}$.  (The absence of magnetic harmonics with $n$ even
follows directly from the antiphase nature of the order.)  Harmonics with $n$
even correspond to charge order.

\section{Temperature dependence of stripe order}

\subsection{ Experimental Results } %\label{Sec:Exp_res}

\begin{figure}
\centerline{\psfig{figure=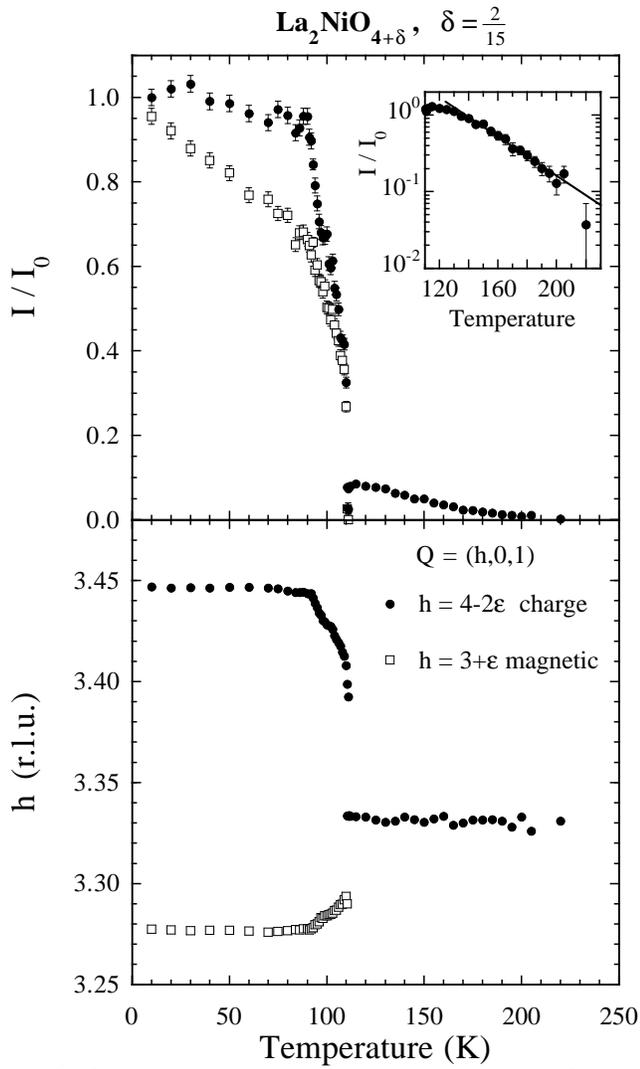,width=3.4in}}
\caption{Temperature dependences of a magnetic peak at $(3+\epsilon,0,1)$  (open
squares) and a charge-order peak at $(4-2\epsilon,0,1)$ (filled circles).  The
upper panel shows integrated intensities normalized with respect to 10~K, the
lower gives the peak position in $h$.  Inset: logarithm of normalized intensity
vs. temperature for the charge-order  peak above 110~K. Line through points is a
linear fit, as discussed in the text.}
\label{fg:Inorm_T_dep}
\end{figure}

The overall temperature dependence of the stripe order is best characterized
by looking at first and second harmonic peaks.  Figure~\ref{fg:Inorm_T_dep}
presents results for a magnetic superlattice peak at 
$(3+\epsilon,0,1)$ and a charge-order peak at $(4-2\epsilon,0,1)$.
The  upper panel  shows the integrated peak intensity normalized with respect to
10~K, while the lower gives the peak position $h$ measured in a $[h,0,0]$ scan.
As found previously,\cite{tran94a,yama94} cooling through $T_m=110.5$~K
results in a discontinuous jump in intensity of both the magnetic and
charge-order peaks, with a concomitant temperature dependence of the peak
positions.  The new feature that we have to report is a peak at
$\epsilon=\frac13$ for $T\ge T_m$. Scattering at this position could come
from either magnetic or charge order (or both); however, the $\bf Q$ 
dependence of the structure factor indicates that it corresponds uniquely to
charge order.  (The intensity of $\epsilon=\frac13$ peaks disappears at small
$Q$, where the magnetic form factor is largest.)

\begin{figure}
\centerline{\psfig{figure=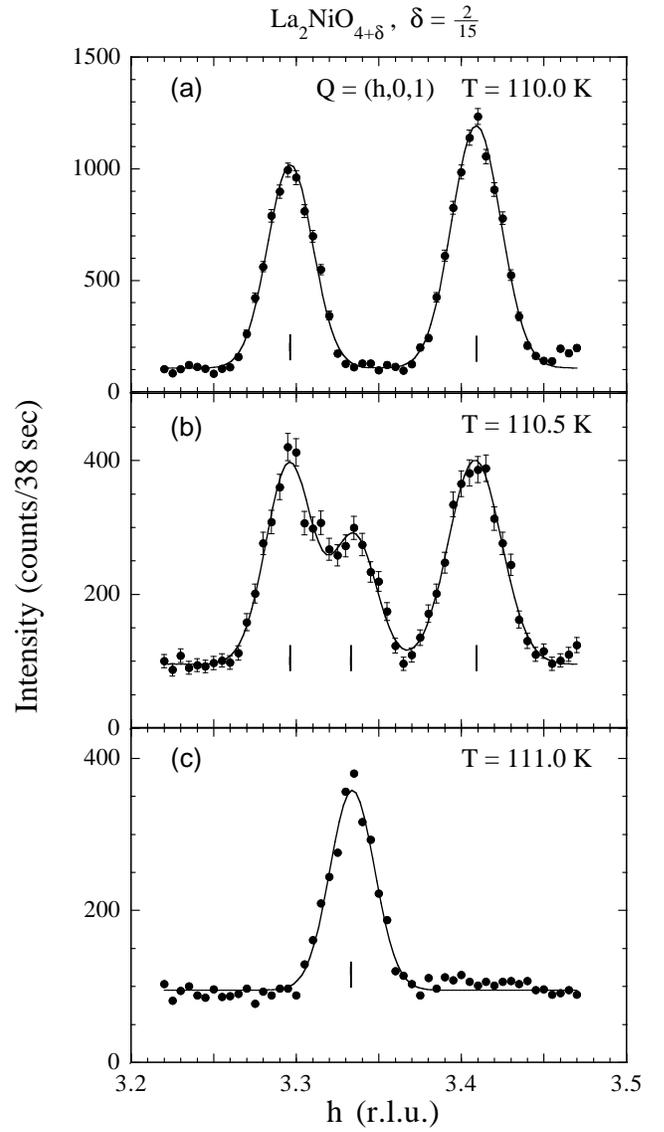,width=3.4in,bbllx=50bp,bblly=50bp,%
bburx=460bp,bbury=770bp}}
\caption{Elastic scans along ${\bf Q}=(h,0,1)$.  Scans were measured at
temperatures of 110~K (a), 110.5~K (b), and 111~K (c).}
\label{fg:elastic_110K}
\end{figure}

The discontinuity of the transition at $T_m$ is illustrated in
Fig.~\ref{fg:elastic_110K}.  At a temperature 0.5~K below the transition we
observe one peak due to magnetic order and another corresponding to charge
order.  Right at $T_m$ a third peak appears at $\epsilon=\frac13$, coexisting
with the previous two.  At 0.5~K above $T_m$ only the $\epsilon=\frac13$
charge-order peak survives.  In contrast to the behavior at $T<T_m$, the
position of the peak remains locked at $\epsilon=\frac13$ while its intensity
gradually decreases with increasing temperature.

The temperature dependence of the $\epsilon=\frac13$ peak is presented on a
logarithmic scale in the inset of Fig.~\ref{fg:Inorm_T_dep}.  The linear
variation of the intensity for $T\gtrsim 150$~K is analogous to a Debye-Waller
like behavior with 
\begin{equation}
 I \sim e^{-2T/T_0},
\label {Eq:Debye-Waller}
\end {equation} and $T_0=67\pm5$~K.  Such a temperature dependence is in sharp
contrast to the usual critical behavior associated with an order-disorder
transition.  In connection with this, it is useful to note that
\begin{equation}
  {\bf g}_{2\epsilon} = \left(\frac23,0,0\right) \equiv 2{\bf g}_{\rm O1},
\end{equation}
so that there is a type of commensurability between the charge stripes in the
planes and the interstitial order along one direction of the lattice.  The
Debye-Waller-like decay of the intensity suggests that the
charge correlations are fluctuating about an average ordered configuration
determined by the ordered interstitial oxygens.  It is possible that the
charge order does not entirely disappear until the interstitials disorder.

\begin{figure}
\centerline{\psfig{figure=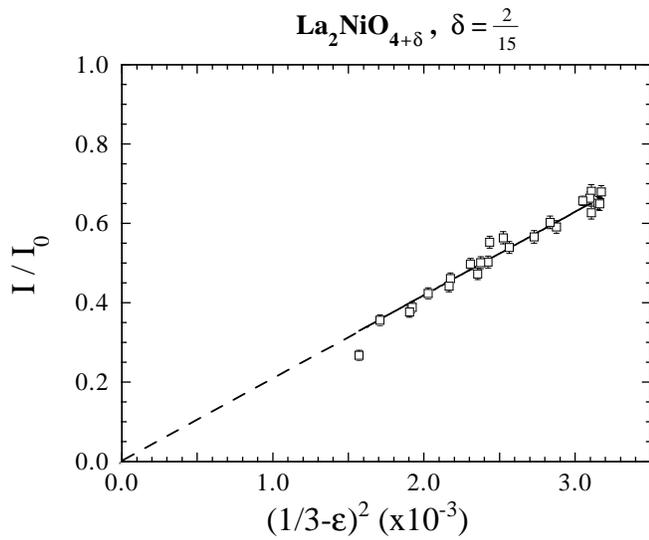,width=3.4in}}
\caption{Normalized intensity of a magnetic peak at $(3+\epsilon,0,1)$ vs. 
$(1/3-\epsilon)^2\sim$ for 80~K$\le T \le T_m$. 
%(b) Normalized intensity of a charge peak at 
%$(4-2\epsilon,0,1)$ vs. $(1/3-\epsilon)^3$. 
Line through points is a linear
fit, as  discussed in the text.}
\label{fg:Inorm_versus_eps}
\end{figure}

The ordering of the spins at $T_m$ is associated with jumps both in the 
charge-order intensity and in $\epsilon$.  Previously\cite{tran94a} it was
found that the superlattice peak intensities show simple power-law correlations
with the quantity
\begin{equation}
  q = \frac 1{3} -\epsilon.
  \label{Eq:q_def}
\end{equation} 
From the measurements between $T_m$ and 80~K, we find for the magnetic-peak
intensity that 
\begin{equation}
  I_{\rm mag} \approx aq^2,
  \label{Eq:q_mag}
\end{equation}
as illustrated in Fig.~\ref{fg:Inorm_versus_eps}.  This result is completely
empirical, as there is no theoretical prediction for the connection between
the intensity and wavevector. 

\begin{figure}
\centerline{\psfig{figure=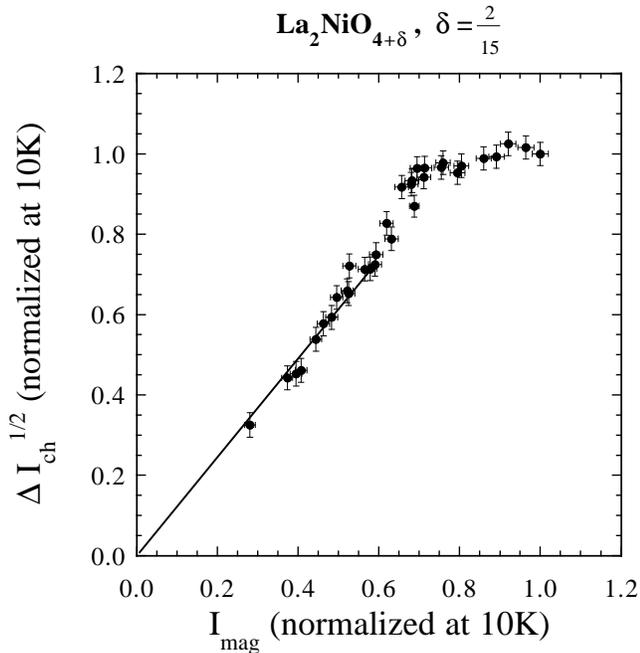,width=3.4in}}
\caption{Intensity of the $(4-2\epsilon,0,1)$ charge-order peak, plotted in the
form $\Delta I_{\rm ch}^{1/2}(T)$ and normalized to the 10-K value, vs. the
normalized $(3+\epsilon,0,1)$ intensity.}
\label{fg:ich_vs_imag}
\end{figure}

A prediction does exist for the relationship between the intensities of the
magnetic and charge order peaks below $T_m$.  Zachar, Kivelson, and
Emery\cite{zach97} have solved an appropriate Landau model for coupled charge-
and spin-density-wave order parameters, $\rho$ and $S$, respectively.  They
find that when the charge is the first to order, it is possible to have a
first-order transition when the spins order, as we observe.  At
temperatures below the magnetic transition, the relationship between the order
parameters is predicted to be, to lowest order,
\begin{equation}
  \rho-\rho_0 = A|S|^2,
  \label {Eq:zachar}
\end{equation} 
where $A$ is a constant and $\rho_0$ denotes the magnitude of $\rho$ just
above the magnetic transition temperature.   Since the intensities of the
magnetic and charge-order peaks are related to the order parameters by
$I_{\rm mag}\sim |S|^2$ and
$I_{\rm ch}\sim |\rho|^2$, Eq.~(\ref{Eq:zachar}) implies that, for $T<T_m$,
\begin{eqnarray}
  \left[I_{\rm ch}(T)\right]^{1/2}-\left[I_{\rm ch}(T_m^+)\right]^{1/2}
  & = & A'I_{\rm mag}(T) \\
  & \equiv & \Delta I_{\rm ch}^{1/2}, \nonumber
  \label{eq:sqrtch}
\end{eqnarray}
where $I_{\rm ch}(T_m^+)$ is the intensity of the charge-order
peak at a temperature just above $T_m$.  Keeping in mind that the model does
not describe the temperature dependence of the wave vector, so that there may
be missing correction terms, we show in Fig.~\ref{fg:ich_vs_imag} that this
relationship is at least plausible near the transition. 

\begin{figure}
\centerline{\psfig{figure=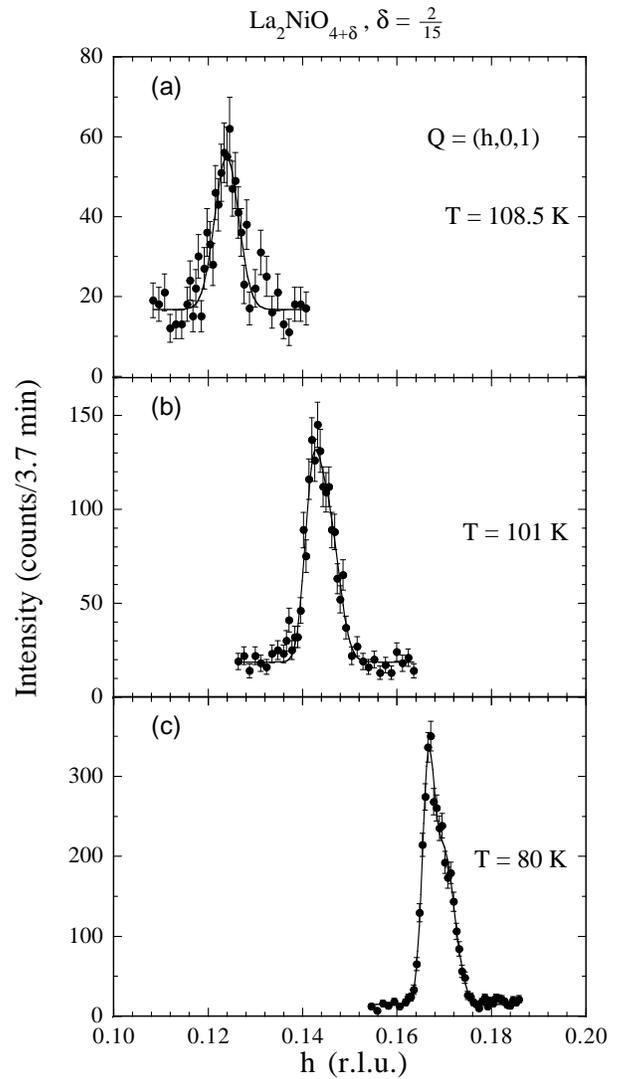,width=3.4in,bbllx=30bp,bblly=50bp,%
bburx=460bp,bbury=770bp}}
\caption{Elastic scans along ${\bf Q}=(h,0,1)$.  Scans were measured at
temperatures of 108.5~K (a), 101~K (b), and 80~K (c).}
\label{fg:elastic_below110K}
\end{figure}

The step-like structures in the lower panel of Fig.~\ref{fg:Inorm_T_dep} are
suggestive of lock-in transitions.  To obtain more precise measurements of
$\epsilon$ versus temperature, we studied a third-harmonic magnetic 
peak, $(1-3\epsilon,0,1)$.  This reflection occurs at a relatively low
scattering angle, and a scan through the peak along the $[h,0,0]$ direction is
nearly transverse to ${\bf Q}$.  For such conditions the resolution in $h$ is
excellent.  In our case the resolution was limited by the sample mosaic
($\approx0.7^\circ$).   At many temperatures the peak shape and width are
limited by the asymmetric mosaic distribution of the sample.  Representative 
scans are shown in Fig.~\ref{fg:elastic_below110K}. At other temperatures we
observed either a single broadened peak, or the coexistence of a broadened
peak and a mosaic-limited one. 

\begin{figure}
\centerline{\psfig{figure=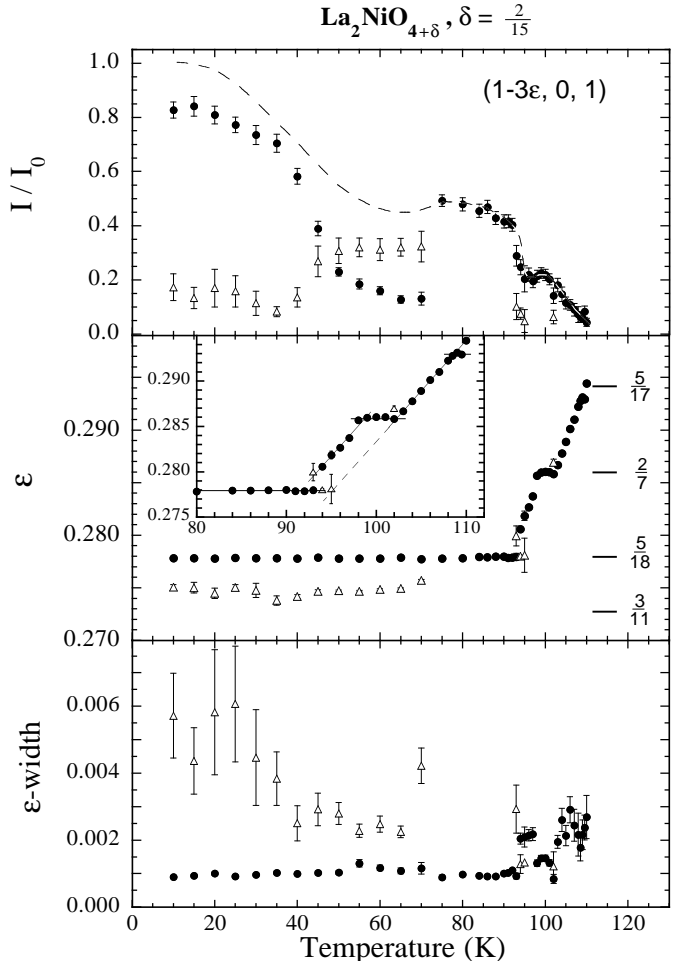,width=3.4in,bbllx=40bp,bblly=50bp,%
bburx=520bp,bbury=770bp}}
\caption{Temperature dependences of third harmonic magnetic peaks at
$(1-3\epsilon,0,1)$. At most temperatures two components with different
$\epsilon$ are coexisting. In all panels,  the (usually, see text) dominant
component is represented by the filled circles and the secondary  by open
triangles. The top panel shows the integrated intensity normalized to the total
intensity from both components at 10~K. The middle panel and inset show
$\epsilon$ and the bottom panel the  peak width, divided by 3 (in units of
$2\pi/a$).}
\label{fg:Tdep_3harm}
\end{figure}

A summary of the results is given in Fig.~\ref{fg:Tdep_3harm}.  
In all panels, the dominant
component is represented by the filled symbols and the  secondary component by
the open ones. (The exception to this rule occurs in the range between 50~K
and 70~K, where the integrated intensity of the broader second component is
actually greater than that of the sharper peak.)  The top panel shows the 
integrated intensity, with the total intensity from both components (dashed
line) normalized to one at 10~K. The middle panel and inset show
$\epsilon$. The bottom panel gives the peak width, divided by 3 (in units of
$2\pi/a$), so that it can be compared directly with the middle panel.

The temperature dependence of $\epsilon$ is rather interesting.  Although it
was measured on warming, it is more convenient to discuss the behavior
starting from $T_m$ and considering the changes as the temperature decreases. 
(Hysteresis effects should be small as long as one cools sufficiently
slowly.)  Initially, $\epsilon$ jumps from $\frac13$ to
$0.2944\approx\frac5{17}$.  From there it decreases linearly with temperature
until it hits a plateau at $0.286=\frac27$.  (There is also a hint of a
plateau at $0.293=\frac{12}{41}$, although the stability of this particular
value is surprising, as discussed below.)  The $\frac27$ plateau extends from
102~K down to 98~K.  Below this, $\epsilon$ jumps to another region of linear
decrease, until it locks into a value of $0.2778=\frac5{18}$ near 93~K.  From
this point down to 10~K a $\frac5{18}$ component is always present; however,
the entire system is locked into this value only down to $\sim75$~K.  Near
70~K the third harmonic splits into two peaks: a sharp one at $\frac5{18}$ and
a broad one centered near 0.275.  The integrated intensity of the broad peak
dominates down to $\sim50$~K, below which most of the intensity shifts back
into the sharp $\frac5{18}$ peak.  In contrast to what was found\cite{tran94a}
with the $\delta=0.125$ crystal, we have not observed any component centered
at $\frac3{11}=0.2727$.

\subsection{ Analysis } \label {Sec:Anal}

In order to make sense of $\epsilon(T)$, we can start with the Zaanen and
Littlewood\cite{zaan94} prediction of charge stripes with a hole density of
one per Ni site.  Assuming that all dopant-induced holes go into O 2p states
lying within the NiO$_2$ layers, one would then expect a domain wall density of
$\epsilon=2\delta$ per row of nickel atoms.  In our case this gives
$\epsilon=\frac4{15}=0.267$.  It appears that the measured $\epsilon$ is
approaching this value as the temperature decreases, but it never quite gets
there.  If we assume that all of the holes are indeed ordered in the stripes,
then the hole density (per Ni site) within a charge stripe, $n_s$, is given by
$n_s=2\delta/\epsilon$.  Even if all holes remain confined to the stripes as
the temperature is raised, the fact that $\epsilon$ increases implies that
$n_s$ is decreasing.  At 80~K we have $n_s=0.960$, at 100~K $n_s=0.933$, and
at $T\gtrsim T_m$ $n_s=0.800$.

The existence of plateaus in $\epsilon(T)$ indicates that the stripe order can
couple to the lattice; however, the almost continuous variation of $\epsilon$
between plateaus shows that the stripe density is not restricted to a few
special values.  To describe the possible stripe arrangements, we need
consider only the one-dimensional ordering along the charge-modulation
direction [see Fig.~\ref{fg:recip_map}(b)].  Within this model, let each row of
Ni or O atoms be represented by a single site.  Thus, there are two Ni sites
per unit cell, separated by $\frac12a$, with an O between each pair of nickels.
The stripes may be pinned to particular sites in the lattice, either by
electron-phonon coupling or Coulomb interaction with the ordered oxygen
interstitials.  The information on the particular pinning sites is contained
in the phase of the scattered neutron waves, which cannot be detected in our
single-scattering measurements.  To be definite, we will assume that a stripe
can be pinned to either a Ni or an O row.  (It will be shown below that this
assumption leads to testable predictions.)  Although it is not necessary for
the present discussion, we will also assume that the charge stripes are quite
narrow.  (The actual charge distribution can be probed by analyzing the
intensities of the higher-harmonic superlattice peaks, as will be discussed in
Sec.~VI.)  For a Ni-centered stripe, we assume  that
there is no static moment on the site where the hole is centered, consistent
with experiment,\cite{tran95b}  but 
neighboring nickels have full spins.  For an O-centered stripe, we assume that
there is no static moment associated with the hole spin.

Since the stripes are charged, they will repel each other.  As a result, the
stripes will arrange themselves so as to maintain the maximum possible
spacing, with the constraint that each stripe is centered on a Ni or O site. 
To understand the possible configurations, we can consider the equivalent
problem of finding the ground state arrangement of integral charges on a one
dimensional lattice at arbitrary  filling, which has been solved by
Hubbard,\cite{hubb78} by Pokrovsky and Uimin,\cite{pokr78} and by Bak and
Bruinsma.\cite{bak82}  For any  given filling, there will be at most two stripe
spacings, which differ by the minimum  increment in stripe position. In our
case, this increment is
$\frac14a$, corresponding to a shift from a Ni to an O site (or vice versa).
The  energy is minimized when the two spacing units are ordered in a periodic
fashion.  The most stable configurations are those with a single spacing.  In
our region of interest, the single-spacing phases occur at 
$\epsilon = \frac14, \frac27$, and $\frac13$, with stripe separations of
$2a,\frac74a$, and $\frac 32a$, respectively.

We can represent these different phases using a notation in which an up or down
arrow represents a Ni spin, a dot represents an O site, and an open circle
represents a stripe. Thus, the  configuration corresponding to $\epsilon =
\frac14$ with Ni-centered stripes looks like 
\begin{eqnarray} \label {Eq:Ni1_4}
 \circ\cdot
 \underbrace{\uparrow\cdot\downarrow\cdot\uparrow}_{\displaystyle{\uparrow}}
 \cdot\circ\cdot
\underbrace{\downarrow\cdot\uparrow\cdot\downarrow}_{\displaystyle{\downarrow}}
 \cdot\circ
\end{eqnarray} 
Note that each spin domain has a net moment, but that the moments
of neighboring domains are antiparallel. Shifting the stripes onto O sites
gives
\begin{eqnarray} 
  \label {Eq:O1_4}
  \uparrow\cdot
  \underbrace{\downarrow\circ\downarrow}_{\displaystyle{\Downarrow}}
  \cdot\uparrow\cdot\downarrow\cdot
  \underbrace{\uparrow\circ\uparrow}_{\displaystyle{\Uparrow}}
  \cdot\downarrow\cdot\uparrow
\end{eqnarray} 
where now we can associate moments with the domain walls.

For $\epsilon = \frac27$, we get equal numbers of Ni-centered and O-centered
stripes:
\begin{eqnarray}
  \label {Eq:Ni_O_2_7}
  \circ\cdot\uparrow\cdot\downarrow\cdot
  \underbrace{\uparrow\circ\uparrow}_{\displaystyle{\Uparrow}}
  \cdot\downarrow\cdot\uparrow\cdot\circ\cdot\downarrow\cdot\uparrow\cdot
  \underbrace{\downarrow\circ\downarrow}_{\displaystyle{\Downarrow}}
  \cdot\uparrow\cdot\downarrow\cdot\circ
\end{eqnarray} 
If we drop the requirement of a single spacing, we can have either
purely Ni-centered stripes:
\begin{eqnarray} 
  \label {Eq:Ni_2_7}
  \circ\cdot
  \underbrace{\uparrow\cdot\downarrow\cdot\uparrow}_{\displaystyle{\uparrow}}
  \cdot
  \circ\cdot\downarrow\cdot\uparrow\cdot\circ\cdot
\underbrace{\downarrow\cdot\uparrow\cdot\downarrow}_{\displaystyle{\downarrow}}
  \cdot\circ\cdot\uparrow\cdot\downarrow\cdot\circ
\end{eqnarray} 
or purely O-centered stripes:
\begin{eqnarray} \label {Eq:O_2_7}
  \underbrace{\downarrow\circ\downarrow}_{\displaystyle{\Downarrow}}
  \cdot\uparrow\cdot\downarrow\cdot
  \underbrace{\uparrow\circ\uparrow}_{\displaystyle{\Uparrow}}
  \cdot\downarrow\cdot
  \underbrace{\uparrow\circ\uparrow}_{\displaystyle{\Uparrow}}
  \cdot\downarrow\cdot\uparrow\cdot
  \underbrace{\downarrow\circ\downarrow}_{\displaystyle{\Downarrow}}
  \cdot\uparrow\cdot\downarrow
\end{eqnarray} 
In the single-spacing phase $\epsilon = \frac13$, Ni-centered stripes give
\begin{eqnarray}
  \circ\,\cdot\uparrow\cdot\downarrow\cdot\circ\cdot\uparrow\cdot\downarrow
  \cdot\circ
\end{eqnarray} 
while O-centered stripes yield
\begin{eqnarray}
  \downarrow\cdot\underbrace{\uparrow\circ\uparrow}_{\displaystyle{\Uparrow}}
  \cdot\downarrow\cdot
  \underbrace{\uparrow\circ\uparrow}_{\displaystyle{\Uparrow}}
  \cdot\downarrow
\end{eqnarray}
For our $\delta=0.133$ crystal, the magnetic order vanishes when $\epsilon$
shifts to $\frac13$; nevertheless, inelastic scattering measurements show that
the spins remain correlated.\cite{tran97c}  In the configuration with O-centered
stripes, the domain walls have moments that are in phase with each other.  As
will be discussed further in Sec.~VI, one might expect the moments of Ni ions
immediately adjacent to an O-centered domain wall to be somewhat reduced
compared to those in the middle of a magnetic domain.  If the modulation of
the moments defining the antiphase domains were purely sinusoidal, then there
would be no net moment in the system; however, any deviation from a pure sine
wave would lead to ferrimagnetism. In contrast, the spins all form
antiparallel pairs when the stripes are centered on Ni rows.  Recently we
demonstrated\cite{tran97b} that one can induce staggered magnetic order in the
$\epsilon=\frac13$ phase by applying a uniform magnetic field.  This result
clearly indicates that the stripes in the high-temperature phase are centered
on O rows.

So far we have considered only phases with uniform stripe spacing, whereas the
real system appears to pass through a range of more complicated phases.  While
we have emphasized the possibility of two types of charge stripes, it is
important to note that the observed values of $\epsilon$ can be explained with
a single type of stripe.  In that case, the relevant phases of uniform spacing
are limited to $\epsilon=\frac14$ and $\frac13$.  To describe the possible
intervening phases it is convenient to introduce a simple notation.  First
note that for $\epsilon=\frac14$, there is 1 spin period in 4 lattice spacings,
or, equivalently, 1 charge stripe per 4 rows of Ni atoms (with 2 rows per
fundamental unit cell).  It follows that this phase can be denoted in terms of
the real-space repeat unit of the charge stripes as $\langle4\rangle$. 
Similarly, $\epsilon=\frac13$ corresponds to $\langle3\rangle$.  The
configuration $\langle43\rangle$ represents alternating stripe spacings of
4 Ni rows and 3 Ni rows.  This corresponds to 2 stripes per 7 Ni rows, or
$\epsilon=\frac27$.  An arbitrary configuration will consist of $n$ 4-row
spacings and $m$ 3-row spacings, giving\cite{tran94a}
\begin{equation}
  \label {Eq:eps_rat_frac}
  \epsilon = \frac{n+m}{4n+3m}.
\end{equation}
For such an arbitrary configuration, the ground-state energy is minimized by
arranging the stripes in a pattern that is as close to uniform as
possible.\cite{hubb78,pokr78}  Some examples are shown in Table~\ref{tab:43}. 
Bak and Bruinsma\cite{bak82} have demonstrated that the possible ground
states form a complete Devil's staircase.

Of course, if two types of stripes are considered, then we must include a
possible stripe spacing of $\frac72$ Ni rows.  Note that to be commensurate
with the Ni-row spacing such units must occur in pairs.  Let the number of
pairs of $\frac72$-spacings per real space repeat unit be $p$.  Then, in the
range $\frac14\le\epsilon\le\frac27$, the values of $\epsilon$ are given by
\begin{eqnarray}
  \epsilon = \frac{n+2p}{4n+7p},
\end{eqnarray}
while, in the range $\frac27\le\epsilon\le\frac13$, the appropriate formula is
\begin{eqnarray}
\epsilon = \frac{2p+m}{7p+3m}.
\end{eqnarray}
Examples are shown in Table~\ref{tab:473}.

One can distinguish between models with either one or two types of stripes by
analyzing the intensities of higher harmonic satellites.  Such an analysis is
discussed in Sec.~VI.  The proposed pinning of the stripes to particular
lattice rows has implications for the layer to layer stacking of the stripes,
as we discuss in the next section.

\section{ Stacking of Stripes } %\label {Sec:Exp:Anal:stack} 

To minimize the Coulomb repulsion between charge stripes in neighboring planes,
the stripes should run parallel to each other with a
staggered alignment.  Such a body-centered type arrangement would maximize the
spacing between nearest neighbors; however, pinning of the stripes to the
lattice would prevent such an ideal staggering. 
Within our model, the stripe pattern should only be able to shift by integral
(or possibly half integral) units of the lattice spacing $a$.  We can
distinguish between these two possibilities, symmetric versus integral-shift
stacking, by analyzing the $l$ dependence of the superlattice intensities.

Consider first the case of a symmetric, body-centered stacking of the stripes.
As discussed in the previous section, a given value of $\epsilon$ can be
written as a ratio of two integers, $r/s$.  The commensurate period of the
magnetic structure is then $sa$, and the period of the charge structure is
$sa/2$.  The displacement of the stripe pattern from one layer to the next is
then $(\frac14s,0,\frac12)$, in lattice units.  The structure factor
describing such a stacking is
\begin{equation}
  F = 1 + e^{i\pi(\frac12sh+l)},
\end{equation}
where the intensity is proportional to $|F|^2$.  For a charge-order reflection
of the type $(2n+2\epsilon,0,l)$, where $n$ is an integer, $|F|^2$ is either 4
or 0.  In the case of $\epsilon=\frac14$, one finds that $|F|^2$ is equal to 0
for $l$ equal to an even integer and 4 for odd $l$, independent of $n$.  For
the general case of $\epsilon=r/s$, however, the value of $|F|^2$ depends not
only on $l$, but also on $r$, $s$, and $n$.  There is a similar variability
for magnetic reflections.  To illustrate this, we have listed values of
$|F|^2$ for a number of different cases in Table~\ref{tab:bc}.  Note in
particular that the body-centered-stacking model implies zero intensity for
the $(4-2\epsilon,0,1)$ reflection when $\epsilon=\frac27$, which is clearly
inconsistent with the results shown in Fig.~\ref{fg:Inorm_T_dep}.
(The tabulated values of $|F|^2$ for $\epsilon=\frac13$ have implications for
the interpretation of magnetic and charge-order scattering as reported in a
recent study\cite{lee97} of La$_{1.67}$Sr$_{0.33}$NiO$_4$.)

Alternatively, if the stripes are pinned to the lattice, then the stripe order
in neighboring layers is described by the basis vectors $(0,0,0)$ and 
$(ma,0,\frac{c}{2})$, where $m$ is an integer 
number.  Taking the simplest case of $m=1$, one finds that, for the $n^{\rm th}$
harmonic superlattice peak,
\begin{equation}
  |F_n|^2 = 2\left[1 + (-1)^l\cos(2\pi n\epsilon)\right].
  \label{eq:Fna}
\end{equation} 
This formula
indicates that, for fixed $h$, the intensities of magnetic peaks should
oscillate as a function of $l$.  It turns out that this is just what is
observed.

\begin{figure}
\centerline{\psfig{figure=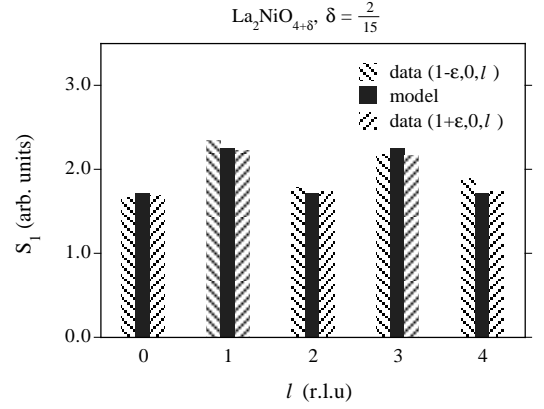,width=3.4in,bbllx=50bp,bblly=60bp,%
bburx=650bp,bbury=500bp}}
\caption{Comparison of the structure factors $S_1(l)$ for the experimental data
(corrected for
$Af^2$) and least-squares fit for the first-harmonic intensities of the
$\frac5{18}$ phase.}
\label{fg:l_depend_1rd}
\end{figure}

We measured integrated intensities for 10 first-harmonic peaks at
$(1\pm\epsilon,0,l)$, with $l=0$ to 4, at the 80~K and 100~K plateaus,
corresponding to $\epsilon=\frac5{18}$ and $\frac27$, respectively. To describe
the experimental results, it is helpful to express the integrated intensities of
the $n=1$ magnetic reflections as
\begin{equation}
  I_1(l) = Af^2|F_1|^2,
\end{equation}
where $A$ is a normalization constant, and we use the parametrization
$f=e^{-Q^2/2\sigma^2}$ to approximate the Ni magnetic form factor as well as the
Debye-Waller factor.  The parameter $\sigma$ was determined by a least-squares
fit to the data.  The values of $|F_1|^2$ extracted from the 80~K measurements
are shown in Fig.~\ref{fg:l_depend_1rd}.  The dominant $l$ dependence of these
results is described by the formula
\begin{equation}
  |F_n|^2 = A'[1+(-1)^l B_n],
\end{equation}
where we have kept the notation general by labelling $F$ and $B$ with $n$, and
the scale factor $A'$ is meant to remind us that we have not determined from
the experimental intensities the absolute magnitude of $|F_1|^2$.  The values
of $B$ characterizing the measurements at 80~K and 100~K are listed in 
Table~\ref{tab:b1234}.  Equation~(\ref{eq:Fna}) predicts that
$B_1=\cos(2\pi\epsilon)$, which yields $B_1=-0.174$ at 80~K
($\epsilon=\frac5{18}$), and $B_1=-0.223$ at 100~K ($\epsilon=\frac27$).  

The model allowing an interlayer offset of the stripe pattern by one lattice
spacing gives qualitative agreement with experiment, although the calculated
magnitude of the parameter $B_1$ appears to be systematically too large by
30\%.  We can get a better fit to the data if we allow for some disorder in
the stacking in terms of different offsets.  The Coulomb energy between charge
stripes in neighboring layers is not the same for all possible integral
offsets.  If we allow for two possible offsets, $a$ and $ma$, with relative
probabilities $1-\alpha$ and $\alpha$, respectively, then, assuming a random
distribution of the offsets, the structure factor becomes
\begin{equation}
  F_n = 1 +\left[(1-\alpha)e^{-i2\pi n\epsilon} + \alpha e^{-i2\pi nm\epsilon}
        \right]e^{-i\pi l}.
  \label{eq:Fnb}
\end{equation}
In the $\frac5{18}$ phase, the Coulomb energy is minimized for $m=1$, and also
for $m=8$.  We fit the measured intensities with this model, varying the
parameters $\alpha$ and $\sigma$ to obtain the results shown in
Fig.~\ref{fg:l_depend_1rd}.  The parameter values and the weighted reliability
factor $R_w$ are listed in Table~\ref{tab:stripe_config}.  For $\frac27$, the
relevant offsets are $m=1$ and 3 (with a shift by $2.5a$ being equivalent in
energy to $m=1$).  Again, the fitting results are listed in
Table~\ref{tab:stripe_config}.  The calculated values of $B_1$ for both cases
are given in Table~\ref{tab:b1234}.

One can see from Eqs.~(\ref{eq:Fna}) and (\ref{eq:Fnb}) that the relative size
of the $l$-dependent intensity oscillation varies with $n$, the harmonic
order.  Thus, another test of our model is to apply it to another value of $n$.
We have parametrized measurements of $(1-3\epsilon,0,l)$ peaks ($l=1$ to 4) in
terms of the coefficient $B_3$, and the results for $T=80$~K and 100~K are
listed in Table~\ref{tab:b1234}.  We then calculated values of $B_3$ using the
parameter values already determined in the fits to the $n=1$ intensities.  The
calculated values are in very good agreement with the measurements.  For
reference, we have also included in Table~\ref{tab:b1234} calculated values of
$B_2$ and $B_4$.

Before leaving this section, it is of interest to consider the consequences of
our model for the $n=2$ charge-order peaks.  Measurements have not detected
any intensity at positions $(2n\pm2\epsilon,0,l)$ with
$l$ even.  Considering just the stacking contribution
(that is, ignoring the structure factor due to the atomic displacement pattern
within a single layer), we can calculate the relative intensities of even-$l$
peaks compared to odd-$l$ peaks.  Using our results for $\epsilon=\frac5{18}$
gives a ratio of 2\%, while for $\epsilon=\frac27$ we obtain 8\%.  These
relative intensities are still quite difficult to detect.  For
$\epsilon=\frac13$ we calculate a ratio of 33\%, which would be detectable
except that the peak position overlaps with the second harmonic of 
${\bf g}_{\rm O1}$.

\section{ Magnetic harmonic satellite spectrum } 
%\label{Sec:anal:harmonic}

Information on deviations from a sinusoidal modulation is contained in the
relative intensities of the higher harmonic superlattice peaks.  In
particular, we are interested in the degree to which the spacing of the
stripes deviates from perfect periodicity as well as the degree to which the
charge is confined within narrow stripes.   By reason of their greater absolute
intensities, the magnetic harmonics are best suited for quantitative analysis.

\begin{figure}
\centerline{\psfig{figure=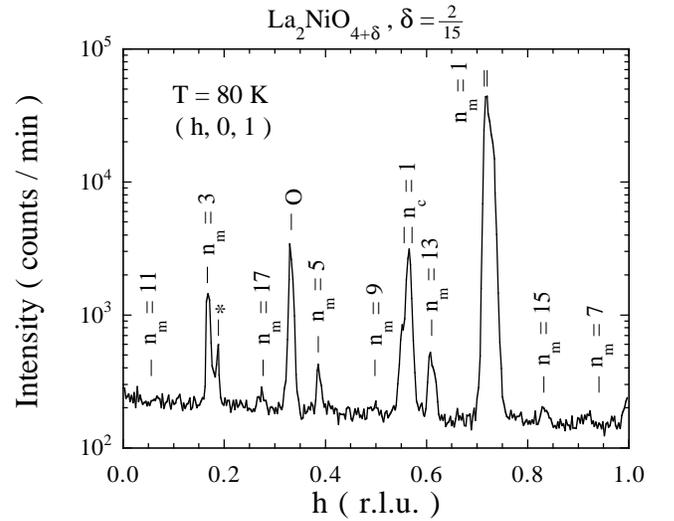,width=3.4in}}
\caption{Spectrum of higher order satellites in the $\frac5{18}$ phase at 80~K
along ${\bf Q}=(h,0,1)$. The $n^{th}$ order magnetic satellites are labeled with
subscript m, the ones due to charge order with c. An oxygen ordering peak is
labeled O and one of undefined origin with (*).}
\label{fg:higher_harm}
\end{figure}

For the $\epsilon=\frac5{18}$ phase found at 80~K, there are 9 independent
harmonics.  Figure~\ref{fg:higher_harm} illustrates the spectrum of higher order
magnetic and charge satellites found in a scan along $(h,0,1)$ for 
$0\le h\le 1$.  Because of the weakness of the higher harmonics, the
intensity is plotted on a logarithmic scale.  The
$n^{th}$ order satellites of magnetic  origin are denoted by the subscript $m$,
while those due to charge order are indicated by the subscript $c$. An oxygen
ordering  peak at $(\frac1{3},0,1)$ (labeled O) and an unexplained peak close
to $n_m=3$ (labeled *) are also recorded.  (The latter very weak peak is
probably due to a misaligned grain.)  Because of insufficient cool-down time for
this special scan, the first magnetic and charge  satellites are split in two
due to partial coexistence of the $\frac5{18}$ and $\frac27$ phases.  Higher
order contributions of the $\frac27$ phase could not be observed in this scan. 

In order to allow a quantitative analysis, separate measurements (following a
more careful cool down) were performed to obtain integrated intensities for all
independent magnetic harmonics.  These were done at 80~K and 100~K to
characterize the $\frac5{18}$ and $\frac27$ phases, respectively.  All peaks
in the range ${\bf Q}=(h,0,1)$ with $-1<h<1$ were measured, and values at $+h$
and $-h$ were averaged.  The results, normalized to the intensity of the first
harmonic, are listed in Tables~\ref{tab:harmonic_5_18} and
\ref{tab:harmonic_2_7}.

To model the harmonic intensities, we must start with a specific model of the
spin density within a unit cell of the stripe pattern, such as the models
shown in Sec.~IV.B.  Within such a one-dimensional unit cell, if we number the
sites with an integer index $m$ and denote the spin on the $m^{\rm th}$ site
as $S_m$, the discrete Fourier transform of the spin arrangement corresponding
to the $n^{\rm th}$ harmonic is then
\begin{equation}
  \tilde{S}_n = \frac1N\sum_{m=0}^{N-1} S_m e^{-i\pi m}e^{-i2\pi n\epsilon m},
\end{equation}
where $N=36$ (14) for $\epsilon=\frac5{18}$ ($\frac27$).  The intensity of the
$n^{\rm th}$-order satellite is given by
\begin{equation}
  I_n\sim |\tilde{S}_n|^2 |F_n|^2,
\end{equation}
where $F_n$ is given by Eq.~(\ref{eq:Fnb}).
Because the measurements span a very small range of $Q$, the variation of the
magnetic form factor and the Debye-Waller factor can be neglected.

We first consider the $\frac27$ case, because it provides the opportunity to
distinguish between three different models: (a) equally-spaced stripes
centered alternately on Ni and O rows, (b) stripes with alternating spacings
centered only on Ni rows, or (c) stripes centered only on O rows.  These three
possibilities are illustrated in Eqs.~(\ref{Eq:Ni_O_2_7})-(\ref{Eq:O_2_7}),
respectively.  In each case, all spins are drawn with the same magnitude
(unity), suggesting quite narrow charge stripes.  To allow for some smoothing
of the spin modulation, we treat the relative magnitude of a spin immediately
adjacent to an O-centered stripe as a fitting parameter.  Since no such sites
occur in model (b), in that case alone we take the magnitude of the spins next
to Ni-centered stripes as the fitting parameter.  The intensities calculated
with each of the models are presented in Table~\ref{tab:harmonic_2_7}, and the
fitted spin in each model is listed in Table~\ref{tab:harmonic_fit_parameter}.
Model (a), with equally-spaced stripes, gives very good agreement with
experiment, and the other two models are ruled out rather convincingly.  As a
further test of model (a), we allowed a second parameter, the size of the spin
adjacent to a Ni-centered stripe, to vary, and found that it decreased only
slightly, from 1.0 to 0.94, while the size of the spin next to an O-centered
stripe remained unchanged.

To model the results for the $\frac5{18}$ phase we considered only a model
similar to (a), with both Ni- and O-centered stripes, although in this case
they are not all equally spaced.  The fit gives the same value as in the
$\frac27$ case for the spin magnitude next to an O-centered stripe (see
Table~\ref{tab:harmonic_fit_parameter}).  The calculated harmonic intensities
are listed in Table~\ref{tab:harmonic_5_18}.  The $R$ factor is somewhat worse
than in the $\frac27$ case, but we also have a substantially larger number of
harmonics to fit.  The quality of the fit seems quite reasonable given the
relative complexity of the structure and the possible systematic errors in
measuring the very weak harmonics.

Given that the analysis of the magnetic harmonics indicates fairly narrow
charge stripes, it is of interest to study the charge-order harmonics.  The
information obtained about the charge modulation is still indirect, because
the neutrons are only directly sensitive to the induced atomic displacements.
Furthermore, quantitative modeling is difficult because of the strong {\bf Q}
dependence of the structure factor, even for a single layer.\cite{tran95b} 
Nevertheless, an attempt was made to detect higher harmonics.  Since the
lowest-order superlattice peaks for charge order correspond to $n=2$, the next
harmonic is $n=4$.  The strongest $n=2$ peak within the accessible $Q$ range
is at $(4-2\epsilon,0,1)$.  Searching in the nearby region we found finite
$n=4$ peaks at $(4-4\epsilon,0,l)$ with $l=0$ and 2.  At $T=8$~K the
intensities of the latter 2 peaks, relative to the strong $n=2$ peak, are
1.4(7)\%\ and 2.2(7)\%, respectively.

\section{Discussion} \label{Sec:Disc}

In the Landau model for coupled charge- and spin-density-wave order
parameters,\cite{zach97} there are two possible ordering scenarios.  If, on
cooling, the first ordering transition is driven by the pure charge term in the
free energy, then spin ordering will tend to occur at a lower temperature;
otherwise, the spin and charge order will appear simultaneously.  Our present
results for the $\delta=0.133$ phase of La$_2$NiO$_{4+\delta}$ show that charge
order appears first, consistent with the behavior found in
La$_{2-x}$Sr$_x$NiO$_4$ (Ref.~\onlinecite{chen93,sach95,tran96a,lee97,rami96})
and in La$_{1.6-x}$Nd$_{0.4}$Sr$_x$CuO$_4$
(Ref.~\onlinecite{tran95a,tran96b}).  Thus, it is the free energy associated
with the charge alone that drives the ordering.  Of course, one must keep in
mind that the charge stripes would not be likely to form and order in the
absence of the magnetic moments of the Ni ions.\cite{zaan94}  
In the charge-ordered phase at
$T>T_m$, the Ni spins remain strongly correlated, as demonstrated by inelastic
neutron scattering measurements reported elsewhere.\cite{tran97c}

The exponential decrease of the charge-order peak intensity at high
temperatures is unusual, but appears to be associated with a commensurability
with the superlattice formed by the interstitial oxygens.  This coupling,
together with the simultaneous appearance of interstitial order and
charge-stripe order as a function of $\delta$, leads one to wonder whether the
energy associated with the charge and spin correlations might influence the
ordering of the interstitial oxygens.  Perhaps this connection could be tested
in a future experiment.

The first-order transition at $T_m$ and the concomitant jump in $I_{\rm ch}$
are consistent with the generic Landau model.\cite{zach97}  The main feature
not explained by that model is the temperature dependence of $\epsilon$.  The
similarity of the series of lock-in transitions to what is observed in
solutions of anisotropic next-nearest-neighbor Ising (ANNNI)
models\cite{bak80,Fish80} suggests the importance of competing interactions. 
(Of course, it is likely that competing interactions are also responsible for
the very existence of stripes in this system.\cite{seul95,kive94,low94,chay96})
One relevant factor is the long-range part of the Coulomb interaction, which
favors a uniform distribution of charge, and hence a large value of
$\epsilon$.\cite{tran96a,kive94,low94,chay96}  A competing factor is the
superexchange energy associated with antiferromagnetic Ni spins.  This
interaction favors wide magnetic domains, and therefore a small value of
$\epsilon$.\cite{kive94,zach97p}  The competition between these energies will
depend on the relative magnitudes of the order parameters $\rho$ and $S$,
resulting in a temperature-dependent $\epsilon$.  The observation\cite{tran96a}
of a very similar temperature dependence (without the lock-ins) in
La$_{2-x}$Sr$_x$NiO$_4$ with $x=0.225$, where the dopant ions are randomly
distributed, demonstrates that this behavior does not require well ordered
dopants.

\begin{figure}
\centerline{\psfig{figure=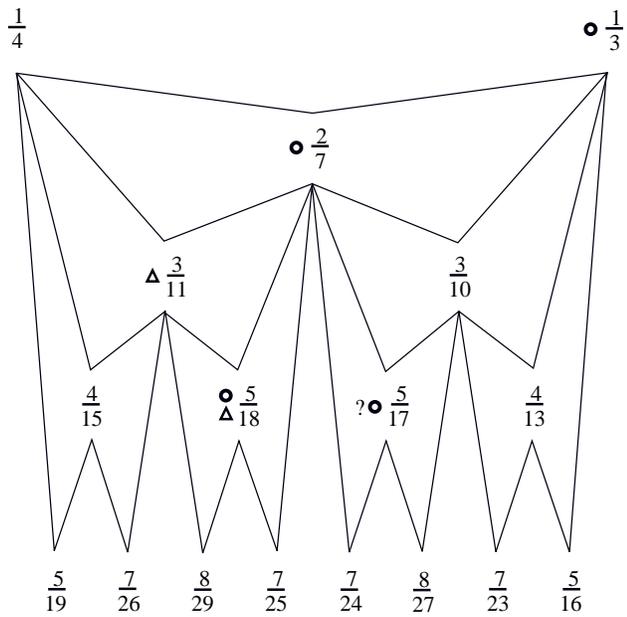,width=3.4in,bbllx=50bp,bblly=170bp,%
bburx=550bp,bbury=650bp}}
\caption{Farey tree composition rule for wave vectors (densities) between 1/4
and 1/3. The observed rational values for $\epsilon$ are marked by the symbols
$\bigcirc$ and $\bigtriangleup$ for La$_2$NiO$_{4.133}$ and 
 La$_2$NiO$_{4.125}$, respectively.}
\label {fg:farey_tree}
\end{figure}

The lock-in transitions provide evidence for some coupling to the lattice.  We
have explained how the values of $\epsilon$ observed at lock-in plateaus can
be understood in terms of long-period commensurate structures.  Only
commensurate structures with relatively short periods can be stabilized at
finite temperature.  Between the plateaus entropy will tend to favor
incommensurate structures.\cite{bak80}  The values of epsilon corresponding to
observed commensurate structures may at first appear to be somewhat arbitrary
rational fractions; however, it is interesting to note that they form the first
levels of a Farey tree.\cite{zach97p,hard54}  A similar correspondence was
observed previously by Shimomura, Hamaya, and Fujii\cite{shim96} in an X-ray
diffraction study of commensurate-incommensurate transitions in the system
[N(CH$_3$)$_4$]$_2$$M$Cl$_4$, where $M=$ Zn, Fe, and Mn.  A Farey tree is
constructed by evaluating a sequence of Farey mediants.  Given two rational
fractions, $n'/n$ and $m'/m$, the Farey mediant is $(n'+m')/(n+m)$.  Starting
with the fractions $\frac14$ and $\frac13$ one obtains the tree shown in
Fig.~\ref{fg:farey_tree}.  The experimentally observed values of $\epsilon$ are
marked.  Farey-tree structures have also been noted in studies of the
two-dimensional Falicov-Kimball model in regions of the phase
diagram dominated by stripe configurations.\cite{wats95}

It was shown previously\cite{tran95b} in a neutron diffraction study of stripe
order in a $\delta=0.125$ sample that the ordered Ni moments are quite large,
with a maximum amplitude that is $>80$\%\ of that observed in undoped
La$_2$NiO$_4$.  Here we have shown that the observed intensities of magnetic
harmonics are consistent with fairly narrow domain walls.  These results,
together with the poor conductivity of the nickelates in general, suggest that
the dopant-induced holes are strongly localized within relatively narrow charge
stripes.  The weakness of the higher harmonics relative to the first harmonic
peak occurs because of the small number of atomic sites between domain walls:
the difference in harmonic intensity between a square wave and a sine wave is
small when there are only a few lattice sites per period.  It follows that the
absence of strong higher harmonics in magnetic scattering studies of
copper-oxide superconductors such as La$_{2-x}$Sr$_x$CuO$_4$ is not
inconsistent with the existence of stripe correlations in those materials.

It is interesting to consider the change in character of the charge stripes
with temperature.  As we have shown elsewhere,\cite{tran97b} the charge
stripes in the $\epsilon=\frac13$ phase, above $T_m$, are centered on oxygen
rows.  From the analysis of the harmonic intensities, we have shown here that,
in the $\epsilon=\frac27$ phase, the charge stripes are half O-centered and
half Ni-centered.  Continuing to lower temperature, the domain walls become
dominantly Ni-centered.  In fact, this shift of the stripes from O- to
Ni-centered is correlated with the evolution of the magnetic order parameter. 
We have observed that, at least near the transition, the intensity of the
first magnetic harmonic is proportional to the square of the quantity $q$
defined by Eq.~(\ref{Eq:q_def}).  One can easily show that $3q$ is equal to the
density of Ni-centered stripes.  The significance of this curious correlation
is not clear.

Along with the shift in lattice alignment,
we have argued that the charge density within the stripes must vary with
temperature.  It follows that the insulating nature of the material cannot be
explained simply by invoking a Peierls transition in 1D half-filled
stripes.\cite{zaan94}  Instead, it appears that the electronic localization
must be understood in terms of interactions in the direction transverse to the
stripes.  While electron-phonon interactions are often discussed, the
temperature dependence of $\epsilon$ makes it clear that purely electronic
effects are more important than coupling to the lattice.  It has been
suggested previously\cite{anis92,loos96} that ``magnetic confinement'' effects
are relatively strong in nickelates, especially when compared to cuprates. 
Perhaps this effect, associated with the size of the transition-metal moments,
is the most important feature distinguishing the nickelates from the cuprates.

\section{ Summary }

We have reported a detailed study of stripe ordering in a single crystal of
La$_2$NiO$_{4+\delta}$ with $\delta=\frac2{15}$.  On cooling, charge order
occurs before magnetic order, although there is a substantial jump in the
charge order parameter at the first-order magnetic transition.  No clear
charge-order transition has been observed, as the intensity of a charge-order
superlattice peak decays exponentially with temperature.  Below $T_m$, the
value of $\epsilon$ decreases with temperature, exhibiting several lock-in
transitions.  A model in which each charge stripe is centered on either a row
of Ni or O ions has been shown to be consistent with the $l$-dependence of the
magnetic peak intensities and with the intensities of higher-order magnetic
harmonic peaks.  Modelling of the latter also indicates relatively narrow
magnetic domain walls.  With decreasing temperature the charge stripes are
initially all centered on O rows, and then begin to shift to Ni rows below
$T_m$, with Ni-centered rows dominating at low temperature.

\acknowledgements

We gratefully acknowledge helpful discussions with V. J. Emery, S. A. Kivelson,
and O. Zachar.  The detailed examination of higher harmonic reflections was
motivated by conversations with P. B. Littlewood.  Enthusiastic encouragement
from J. Zaanen is appreciated.  Work at Brookhaven was carried out under
Contract No.\ DE-AC02-76CH00016, Division of Materials Sciences, U.S.
Department of Energy.

%\bibliographystyle{prsty}
%\bibliography{lno,theory}

%       %       %       figure captions         %       %       %

%       %       %       tables          %       %       %

\onecolumn
\renewcommand{\arraystretch}{1.2}

\newpage
\narrowtext

\begin{table}[tbp]
\caption{Some of the more stable stripe configurations in the range $\frac14 <
\epsilon < \frac13$ assuming a single type of charge stripe. The notation is
explained in the text. }
\label{tab:43}
\begin{tabular}{lccll}                                     
$n$ & $m$ &\multicolumn{2}{c}{$\epsilon$} & Configuration \\
 &&fraction & decimal \\   
\hline
2 & 1 & $\frac3{11}$  & 0.2727  & $\langle443\rangle$ \\  
5 & 3 & $\frac8{29}$  &  0.2759 & $\langle44344343\rangle$ \\  
3 & 2 & $\frac5{18}$  & 0.2778 & $\langle44343\rangle$ \\  
1 & 1 & $\frac27$ & 0.2857 & $\langle43\rangle$ \\
2 & 3 & $\frac5{17}$ & 0.2941 & $\langle43343\rangle$ \\ 
\end{tabular}
\end{table}

\begin{table}[tbp]
\caption{Some of the more stable stripe configurations in the range $\frac14 <
\epsilon < \frac13$ assuming two types of stripes. The notation is explained in
the text. }
\label{tab:473}
\begin{tabular}{lcccll}                                     
$n$ & $p$ & $m$ & \multicolumn{2}{c}{$\epsilon$} & Configuration \\
 &&&fraction & decimal \\   
\hline
1 & 1 & 0 & $\frac3{11}$  & 0.2727 &
 $\left\langle4\frac72\frac72\right\rangle$ \\ 
2 & 3 & 0 & $\frac8{29}$  & 0.2759 &
 $\left\langle4\frac72\frac72\frac72 4\frac72\frac72\frac72\right\rangle$ \\
1 & 2 & 0 & $\frac5{18}$  & 0.2778 & 
 $\left\langle4\frac72\frac72\frac72\frac72\right\rangle$ \\  
0 & 1 & 0 & $\frac27$  &  0.2857 &
 $\left\langle\frac72\frac72\right\rangle$ \\
0 & 2 & 1 & $\frac5{17}$ & 0.2941 &
 $\left\langle\frac72\frac72\frac72\frac72 3\right\rangle$ \\ 
\end{tabular}
\end{table}

\mediumtext

\begin{table}[tbp]
\caption{Examples of $|F|^2$ for both magnetic and charge-order peaks in the
case of body-centered stacking of stripe-ordered layers.}
\label{tab:bc}
\begin{tabular}{ccccccccc}                                     
$\epsilon$ & \multicolumn{2}{c}{$|F(1\pm\epsilon,0,l)|^2$} &
\multicolumn{2}{c}{$|F(3\pm\epsilon,0,l)|^2$} &
\multicolumn{2}{c}{$|F(2\pm2\epsilon,0,l)|^2$} &
\multicolumn{2}{c}{$|F(4\pm2\epsilon,0,l)|^2$} \\
 & $l$ even & $l$ odd & $l$ even & $l$ odd & $l$ even & $l$ odd & $l$ even &
$l$ odd  \\   
\hline
$\frac14$    & 2 & 2 & 2 & 2 & 0 & 4 & 0 & 4 \\
$\frac3{11}$ & 0 & 4 & 4 & 0 & 4 & 0 & 0 & 4 \\
$\frac5{18}$ & 2 & 2 & 2 & 2 & 0 & 4 & 0 & 4 \\
$\frac27$    & 2 & 2 & 2 & 2 & 0 & 4 & 4 & 0 \\
$\frac13$    & 4 & 0 & 0 & 4 & 4 & 0 & 0 & 4 
\end{tabular}
\end{table}

\narrowtext

\begin{table}[tbp]
\caption{Values of the $l$-dependent coefficient $B_n$ (described in the text)
for harmonics
$n=1$--4. For $n=1$, experimental and fitted values are given.  For $n=3$,
experimental values are compared with values calculated using the parameters
determined from the $n=1$ fit.  Only calculated values are given for $n=2$ and
4.}
\label{tab:b1234}
\begin{tabular}{ccccccc}                                     
$T$ & \multicolumn{2}{c}{$B_1$} & \multicolumn{2}{c}{$B_3$} & $B_2$ & $B_4$ \\
(K) & expt. & fit & expt. & calc. & calc. & calc. \\   
\hline
 80 & $-0.13(1)$ & $-0.132$ & 0.34(7) & 0.398 & $-0.964$ & 0.841 \\
100 & $-0.15(1)$ & $-0.153$ & 0.54(2) & 0.508 & $-0.854$ & 0.508 
\end{tabular}
\end{table}

\begin{table}[tbp]
\caption{Fitted values of the occupation fraction $\alpha$ for layers with basis
vector $(ma,0,\frac{c}{2})$ in the average magnetic unit cell and $\sigma$,
the Ni magnetic  form factor constant. \(R_w=\sum w_n^2 (I_{obs}-I_n)^2 / \sum
w_n^2 I_{obs}^2;\;  w_n=\) statistical weight). }
\label {tab:stripe_config}
\begin{tabular}{rclllllll}                                    
\multicolumn{1}{c}{T}&\multicolumn{1}{c}{$\epsilon$}&\multicolumn{1}{c}{$m$} &
\multicolumn{1}{c}{$\alpha$} &
\multicolumn{1}{c}{$\sigma$} & \multicolumn{1}{c}{$R_w$}\\
\hline 
80~K   & $\frac5{18}$ &  8 &  0.123(2) & 3.2(2) & 0.034 \\  
100~K   & $\frac27$ &  3 &  0.1182(2)  &3.1(2) & 0.035 \\ 
\end{tabular}

\end{table}

%\begin{table}[tbp]
%\caption{Comparison of the l-dependence for the observed values of $|F_3|^2$ of
%the $3^{rd}$ order harmonic for $(1-3\epsilon,0,1), 
%\epsilon=\frac5{18}$,  with those calculated using the parameters determined in
%a fit to the $1^{\rm st}$ order magnetic satellite.}
%\label{tab:l_depend_fit}
%\begin{tabular}{rll}                                     
%$l$ & $|F_3|^2_{\rm obs} $ & $|F_3|^2_{\rm calc}$ \\   
%\hline 1   & 1.65(11) &  1.13   \\  2   & 3.00(24)  &  2.67   \\  3   &
%1.31(35) &  1.13   \\  4   & 3.35(57)  &  2.67   \\ 
%\end{tabular}

%\end{table}

\begin{table}[tbp]
\caption{Relative intensities of the $n^{th}$ order magnetic satellite peaks
in the $Q$ range $(h,0,1)$ with $-1<h<1$ for $\epsilon=\frac5{18}$.  Both
observed  and fitted values are listed.}
\label{tab:harmonic_5_18}
\begin{tabular}{rll}                                     
$n$ & $(I_n/I_1)_{\rm obs}$ (\%) & $(I_n/I_1)_{\rm calc}$ (\%) \\   
\hline
3   & 1.70(21)  & 1.008 \\  
5   & 0.10(10)  & 0.168 \\  
7   & 0.10(7)   & 0.024 \\  
9   & 0.001(1)  & 0.136 \\ 
11  & 0.001(1)  & 0.014 \\ 
13  & 0.90(16)  & 0.642 \\ 
15  & 0.10(3)   & 0.309 \\ 
17  & 0.10(5)   & 0.121  \\
\hline 
$R_w$   &  & 0.47  
\end{tabular}

\end{table}

\begin{table}[tbp]
\caption{Same as previous table, but for $\epsilon=\frac27$.  The different
models are described in the text.}
\label {tab:harmonic_2_7}
\begin{tabular}{lllll}                                     
$n$ & $(I_n/I_1)_{\rm obs}$ (\%) & 
\multicolumn{3}{c}{$(I_n/I_1)_{\rm calc}$ (\%)} \\
 & & model (a) & model (b) & model (c) \\
\hline 
3   & 0.81(19) &  0.910  &   0.355  & 1.836 \\  
5   & 0.04(4)  &  0.051 &   0.063  & 0.217 \\ 
\hline 
$R_w$   &  & 0.13  &   0.56 & 1.42
\end{tabular}

\end{table}

%\begin{table}[tbp]
%\caption{Values for the magnitude of the Ni-spin close to a domain wall
%obtained in fitting  the intensity ratios of the $n^{th}$ order versus first
%order magnetic satellite peak for 
%$(h+n\epsilon,0,1)$. In model $a$ for $\epsilon=\frac27$, the only fit
%parameter is the magnitude of the  Ni-spin close to the oxygen centered domain
%wall. Model
%$b$ contains only Ni centered domain walls and
%$c$ only O-centered domain walls. }
%\label{tab:harmonic_fit_parameter}
%\begin{tabular}{llll}                                     
%\multicolumn{3}{c}{$\epsilon=\frac2{7}$} &
%\multicolumn{1}{c}{$\epsilon=\frac5{18}$} \\
% model (a) & model (b) & model (c) \\% & O-centered \\   
%\hline
% 0.65(2) &  0.44(4)   & 0.26(8) & 0.66(6)   \\ 
%\end{tabular}

%\end{table}

\begin{table}[tbp]
\caption{Values for the relative magnitude of the Ni spins adjacent to a domain
wall obtained in fitting the relative intensities of the higher order
magnetic satellite peaks.  Models (a), (b), and (c) are described in
the text.}
\label{tab:harmonic_fit_parameter}
\begin{tabular}{cccc}                                     
model &  \multicolumn{3}{c}{Fitted reduction factor for %
 $\langle S_{\rm Ni}\rangle$ near DW} \\
 & & $\epsilon=\frac27$ & $\epsilon=\frac5{18}$ \\   
\hline
(a) & & 0.65(2) & 0.66(6) \\ 
(b) & & 0.44(4) \\
(c) & & 0.26(8)
\end{tabular}

\end{table}

\end{document}